# Constraining variability of coupling constants with bright and extreme quasars

Rajendra P. Gupta*
*Department of Physics, University of Ottawa, Ottawa, Canada K1N 6N5*

**ABSTRACT**

The 'extreme Population A (xA) quasars' approaching, sometimes exceeding the Eddington limit, are a type of quasars that could serve as standard candles to measure distances too large for supernovae type Ia (SNe Ia) to be observable. For using xA quasars as standard candles, it would be beneficial to know how their luminosities would vary if coupling constants varied over cosmic time. Alternatively, when calibrated using SN Ia standard candle, xA quasar observations could constrain the variation of coupling constants. We show that the Hubble diagram of xA quasars provides the same constraint on the constants' variation as the Hubble diagram of SNe Ia from the Pantheon data. The coupling constants vary concurrently in our model, i.e., the variation of the speed of light $c$, the gravitational constant $G$, the Planck constant $h$, and the Boltzmann constant $k$, are interrelated as $G \sim c^3 \sim h^3 \sim k^{3/2}$. The constraint thus determined can be expressed in terms of the Hubble constant $H_0$ as $\left(\dot{G}/G\right)_0 = 3(\dot{c}/c)_0 = 3\left(\dot{h}/h\right)_0 = 3/2\left(\dot{k}/k\right)_0 = 5.4 H_0 = 3.90(\pm 0.04) \times 10^{-10} \text{ yr}^{-1}$, where subscript 0 is for the current time. This conclusion is corroborated with the analysis of data on the quasars with correlated UV and X-ray emissions.

## 1. INTRODUCTION

Quasars are the brightest non-transient objects in the Universe and thus are observable even at high redshifts representing the epoch when galaxies were first formed. The very high luminosity of these compact sources results from the gravitational energy released from the accretion of surrounding matter on supermassive black holes. The very high luminosity generates radiation pressure on accreting mass, limiting the rate of mass accretion (Eddington limit), which limits the quasar luminosity (Eddington luminosity). Quasars have been observed at redshifts up to $z = 7$, two of them even at $z > 7$ (Mortlock et al 2011, Banados et al 2018). The one observed at $z = 7.53$ has the bolometric luminosity of $4 \times 10^{13}$ solar luminosity and a black-hole mass of $8 \times 10^8$ solar masses. This quasar came into existence about 690 million years after the big-bang, the epoch when the reionization of hydrogen in the Universe was not complete. One could thus infer that the massive black hole grew early (Latif et al. 2013, Alexander & Natarajan 2014) and possibly accreted matter at hyper-Eddington accretion rate (Pacucci et al. 2015, Inayoshi et al. 2016). So, it is natural to explore the possibility of using them as standard candles, especially beyond the range of supernovae type Ia (SNe Ia) standard candles. However, while a supernova's luminosity is easy to establish from its time-dependent spectral features, it is not the case for quasars. The challenge is to define and measure standard luminosity in order to employ quasars to measure distances. Our primary interest is to explore how quasars evolve in a variable physical constants (VPC) universe and if they could be used to constrain the variation of the constants.

Many approaches have been explored to reliably determine quasars' luminosity from observations. Among them are (i) the relationship between the continuum and the emission-line fluxes (Baldwin, 1977, Korista et al. 1998), (ii) relation between the distribution of the emission spectrum linewidths, e.g., C IV in NGC 4151, and the inclination of the quasar's broad line region (BLR) and luminosity function (Fahey et al 1991, Rudge & Raine 1999), (iii) the variability of time delay among various wavelength regions of the emission spectrum (Collier et al 1999), (iv) the empirical relationship between the BLR radius and the luminosity (Watson et al 2011, Kilerci Eser et al 2015), (v) properties of highly accreting quasars approaching Eddington limit (Wang et al 2013, Marziani and Sulentic 2014, Dultzin et al 2020), (vi) variability of X-ray emission with the mass of the supermassive black hole (La Franca et al 2014), (vii) reverberation mapping for determining apparent size of the BLR region and related luminosity (Blandford & McKee 1982, Peterson 1993, Elvis & Karovska 2002), and (viii) non-linear relation between UV and X-ray emission (Lusso & Risalti 2017, Risaliti & Lusso 2019, Lusso et al 2020, Bisogni et al 2021). Here we will consider two approaches from the above eight that are well developed and most current, i.e., v and viii.

Quest for constraining the gravitational constant has continued unabated since Dirac (1937) suggested it to be evolutionary based on his large number hypothesis. Teller (1948) considered how the variation of $G$ will affect the luminosity of Sun, and hence the life on Earth in the past, and thereby proposed the limits on $G$ variation. Constraints on $\dot{G}/G$ well below that predicted by Dirac have also been determined by various other methods developed since then. The methods include: (i) solar evolution (e.g., Sahini & Shtanov 2014); (ii) lunar occultation and eclipses (Morrison 1973); (iii) paleontological evidence (Sisterna & Vucetich); (iv) white dwarf cooling and pulsation (e.g., Corsico et al. 2013); (v) star cluster evolution (Degl'Innocenti 1996); (vi) neutron star masses and ages (Thorsett 1996); (vii) CMB anisotropies (e.g., Ooba et al. 2017); (viii) big-bang nucleosynthesis abundances (e.g., Alvey et al. 2020); (ix) asteroseismology (Bellinger & Christensen-Dalsgaard 2019); (x) lunar laser ranging (e.g., Hofmann and Müller 2018); (xi) the evolution of planetary orbits (e.g., Genova et al. 2018); (xii) binary pulsars (e.g., Zhu et al. 2019); (xiii) supernovae type-1a (SNeIa) luminosity evolution (e.g., Wright & Li 2018); and (xiv) gravitational-wave observations on binary neutron stars (Vijaykumar et al. 2021).

*email: rgupta4@uottawa.ca

Einstein (1907) considered a possible *variation* of the speed of light $c$, even though his celebrated theory of relativity is founded on the basis of the *constancy* $c$. Using the observational data of supernovae 1a (SNe1a), baryon acoustic oscillations (BAO), Hubble parameter $H(z)$, cosmic microwave background, and power-law $c$ variation for very low and moderate redshift $z$ values, Qi et al. (2014) reported negligible $c$ variation. Another possibility of measuring the temporal variation of $c$ was suggested by Salzano et al. (2015). From the relation between the maximum value of the angular diameter distance $D_A(z)$ and $H(z)$, they considered determining constraints on the variation of $c$ by using the BAO and simulated data. The independent determination of Suzuki et al. (2012) of $H(z)$ and luminosity distance $D_L(z)$ from SNe1a observations was used by Cai et al. (2016) to examine the variation of $c$. Cao et al. (2017) reported the first measurement of $c$ value with respect to $z = 1.7$ and found it essentially the same as measured at $z = 0$, i.e., on Earth, by using the angular diameter distance measurement for radio quasars extending to high redshifts. A direct determination of $c$ variation using the galactic-scale measurements of strong gravitational lensing systems with SNe1a and quasars as the background sources was proposed by Cao et al. (2020). By the statistical analysis of a galaxy-scale strong gravitational lensing sample, including 161 systems with stellar velocity dispersion measurements, Lee (2021) showed essentially no variation in the speed of light. Mendonca et al. (2021) estimated the variation of $c$ using galaxy cluster gas mass fraction measurements with negative results.

The Planck constant $h$ and the Boltzmann constant $k$ are other constants of interest in our work. The impact of time-dependent stochastic fluctuations of Planck constant was studied by Mangano et al. (2015). The effect of a varying Planck constant on mixed quantum states was studied by de Gosson (2017). Temporal and spatial variation of the Planck constant by elevating it to the dynamical field that couples to other fields and itself through the Lagrangian density derivative terms and the cosmological implications of such variations was studied by Dannenberg (2018), who also reviewed the subject literature. The Doppler broadening of absorption lines in thermal equilibrium (e.g., Castrillo et al. 2009, Djerroud et al. 2009), such as the profile of rovibrational line of ammonia along a laser beam, is the basis of direct measuremenyt of the Boltzmann constant, since such profile is associated with the Maxwell-Boltzmann molecular velocity distribution through the kinetic energy contained in each molecule. Thus, spectral line profiles of distant objects, such as quasars and interstellar media, should make it possible in principle to constrain the variation of Boltzmann constant.

Comprehensive reviews on the variation of fundamental constants have been done by Uzan (2003, 2011). Concerns about the validity of the variability of dimensioned vs. dimensionless constants are debated in these reviews and elsewhere in the literature (e.g., Ellis & Uzan 2005, Duff 2002, 2014).

One normally studies the potential variation of one constant while assuming all others pegged to their current value. When several constants in the interpretation of the data thus collected involve multiple constants, it is not a prudent approach albeit necessary to control the number of unknown variables. Nevertheless, if the constants' variations are interrelated, e.g., $G \sim c^3 \sim h^3 \sim k^{3/2}$, it becomes possible to permit concurrent variation of $c$, $G$, $h$, and $k$. The interrelationaship was derived from our cosmological, astrophysical and astrometric studies in the course of: (i) resolving the primordial lithium problem (Gupta 2021a), (ii) finding a reasonable solution to the faint young Sun problem (Gupta 2022a), (iii) showing that orbital timing studies do not constrain the variation of $G$ (Gupta 2021b), (iv) proving that gravitational lensing cannot determine the variation of $c$ (Gupta 2021c), and (v) establishing that SNe Ia data is consistent with the VPC model (2022b).

We examine the basic physics of xA quasars and study how it is modified for the VPC model in Section 2. Section 3 is for deriving expressions for the luminosity distance and the distance modulus. In Section 4, we take the Hubble diagram data of Dultzin et al. (2020) and see how our model fits it. Section 5 is to determine the function of the evolution of the black hole mass with the redshift for xA type quasars and compare it with that determined from fitting the Hubble diagram data. In Section 6, we attempt to study the type of quasars whose X-ray and UV emissions are correlated, following Lusso and Risaliti (2017) and Risaliti and Lusso (2019), and we study their Hubble diagram in Section 7. We discuss our findings in Sections 8 and summarize our conclusion in Section 9.

## 2. EXTREME POPULATION A QUASARS

The 'extreme Population A (xA) quasars' approaching, sometimes exceeding the Eddington limit, are a class of quasars that could serve as standard candles to measure distances too large for supernovae type Ia to be observable (e.g., Marziani et al. 2003a; Teerikorpi 2005; Bartelmann et al. 2009; Sulentic, Marziani & D'Onofrio 2012; Wang et al. 2013; La Franca et al. 2014, Marziani & Sulentic 2014, Dultzin et al. 2020).

Three conditions need to be satisfied for the possible use of xA quasars as standard candles (Dultzin et al. 2020):
1. Eddington ratio $L/L_E \equiv \lambda_E$ is constant where $L$ is the luminosity of the quasar and $L_E$ is its Eddington luminosity.

2. The black hole mass can be expressed with the virial relation $M = r\delta v^2/G$. Here $r$ is the radius of the broad-line region (BLR) of the emitted radiation, $\delta v$ is the virial velocity in the region, and $G$ is the gravitational constant.
3. They should have spectral invariance, i.e., the ionization parameter $U = Q(H)/(4\pi r^2 n_H c)$ should be constant. Here $Q(H)$ is the number of hydrogen ionizing photons, $n_H$ is the hydrogen number density, and $c$ is the speed of light.

We will explore if the above conditions are affected under the varying coupling constants scenario. If the conditions are affected, could they constrain the constants' variation?

The luminosity $L$ of a quasar with a black hole of mass $M$ that is accreting mass at a rate $\dot{M}$ from a disk of the inner diameter $r_{in}$ and outer diameter $r_{out}$ with $r = r_{in} \ll r_{out}$ is given by (Maoz 2016)

$$L = \frac{1}{2}\left(\frac{GM\dot{M}}{r_{in}}\right), \tag{1}$$

whereas the Eddington luminosity $L_E$ is

$$L_E = \frac{4\pi c G M m_p}{\sigma_T}. \tag{2}$$

Here $m_p$ is the proton mass and $\sigma_T$ is the Thomson scattering cross-section representing the scattering of photons by electrons. Since $L_E$ is the limiting luminosity, $L \leq L_E$. Therefore, from equations (1) and (2)

$$\dot{M} \leq 8\pi r c m_p/\sigma_T. \tag{3}$$

Let us now see how the above three expressions scale when the coupling constants vary as follows (Gupta 2022b, 2022c):

$$c = c_0 g(a); G = G_0 g(a)^3; h = h_0 g(a); k_B = k_{B,0} g(a)^2 \tag{4}$$

$$G \sim c^3 \sim h^3 \sim k_B^{3/2} \sim g^3, \text{ and } \frac{\dot{G}}{G} = 3\frac{\dot{c}}{c} = 3\frac{\dot{h}}{h} = \frac{3}{2}\frac{\dot{k}_B}{k_B}. \tag{5}$$

Here the subscript 0 refers to the current time, $h$ is the Planck constant, and $k_B$ is the Boltzmann constant with $g(a)$ the function relating the variation of the constants with the cosmological scale factor $a \equiv 1/(1+z)$, with $z$ as the redshift. These relations are independent of the form of $g(a)$, i.e., $g(z)$. However, the form we have successfully used in the past (where we have written $g$ as $f$) is

$$g = \exp(a^\alpha - 1) \equiv \exp[(1+z)^{-\alpha} - 1]. \tag{6}$$

Here $\alpha$ is the parameter representing the strength of the variation of the constants. We have found $\alpha = 1.8$ analytically (Gupta 2018) and confirmed it from the analysis of various observations (Gupta 2020, 2021a, 2021b, 2021c, 2022a, 2022b).

We may now write the scaling of the expressions (equations 1 – 3), realizing that any distance is measured using the speed of light, i.e., $r \sim c \sim g$. However, we should first see how the Thomson scattering cross-section $\sigma_T$ scales. It is given by

$$\sigma_T = \frac{8\pi}{3}\left(\frac{q^2}{4\pi\epsilon_0 mc^2}\right)^2. \tag{7}$$

Here $q$ is the charge and $m$ is the mass of the particle, and $\epsilon_0$ is the permittivity of space that is related to the speed of light through $c = 1/\sqrt{\epsilon_0 \mu_0}$ where $\mu_0 = 4\pi \times 10^7$ H m$^{-1}$ is the permeability of space. Thus, $\epsilon_0 \propto 1/c^2$. Since charge and mass are considered invariant when coupling constants vary, we get $\sigma_T \sim g^0$. Therefore,

$$\dot{M} \leq 8\pi r c m_p/\sigma_T \sim g^2, \tag{8}$$

$$L = \frac{1}{2}\left(\frac{GM\dot{M}}{r_{in}}\right) \sim \frac{g^3 \dot{M}}{g} \sim g^4, \tag{9}$$

$$L_E = \frac{4\pi c G M m_p}{\sigma_T} \sim g^4. \tag{10}$$

Thus, the Eddington ratio, $L/L_E \sim g^0$, is unaffected by the variation of the coupling constants: One cannot use it for constraining the variation of the constants.

Coming now to the black hole mass determination, we may write again

$$M = r\delta v^2/G. \tag{11}$$

Since $r \sim g$, and velocity is time derivative of length measured with the speed of light $c$, it also scales as $g$. Therefore, the mass of the black hole, $M \sim g g^2/g^3 \sim g^0$, is unaffected by the constants' variation: We cannot consider it for constraining the constants.

Next, we should consider the spectral invariance, which may be rewritten as

$$U = \frac{Q(H)}{4\pi r^2 n_H c}. \tag{12}$$

The hydrogen number density is inversely proportional to volume, i.e., $n_H \propto 1/length^3 \sim g^{-3}$. Therefore, $\sim Q(H)/(g^2 g^{-3} g) \sim Q(H) g^0$. But $Q(H)$ represents the number of hydrogen ionizing photons, which does not change with $z$ under the VPC scenario. While the ionizing photon energy evolves with $z$, it is exactly offset by the evolution of the energy required to ionize hydrogen atoms (Gupta 2022b). Thus, $U \sim g^0$, i.e., $U$ does not vary on account of the varying coupling constants, and therefore cannot be used for constraining them.

### 3. LUMINOSITY DISTANCE AND DISTANCE MODULUS

We will use the same VPC approach to fit the quasar data we used to fit the SNe Ia data (Gupta 2022b). However, the approach will be modified since, among other things, quasar luminosities depend strongly on their black hole masses that are known to increase with the redshift (e.g., Vestergaard & Osler 2009).

Ordinarily, the flux $F$ received from a distant source of standard luminosity $L_s$ at proper distance $d_P$ with redshift $z$ may be written as (Ryden 2017)

$$F = \frac{L_s}{4\pi S_k(d_P)^2} \frac{1}{(1+z)^2}. \tag{13}$$

The expression includes the flux reduction due to expansion of the Universe that causes the energy of the photons to decrease by $1/(1+z)$ and due to time dilation by another factor $1/(1+z)$. The curvature of the Universe is expressed through $S_k(d_P) = R\sin(d_P/R)$ for $k = +1$ (closed Universe); $S_k(d_P) = d_P$ for $k = 0$ (flat Universe), $S_k(d_P) = R\sinh(d_P/R)$ for $k = -1$ (open Universe), where $R$ is the parameter related to the curvature of the Universe.

Equation (13) is modified when the coupling constants are evolutionary. The time dilation due to changing speed of light $c = c_0 g(z)$ modifies the flux by an additional factor $1/g(z)$ (Gupta 2022b). Now from equation (9) we may write the luminosity in the VPC universe $L_v = L_s g(z)^4$, i.e., $L_s = L_v g(z)^{-4}$.

It is well known that the mass of the quasars increases with $z$ (e.g., Vestergaard & Osmer 2009), and as a result, its luminosity (equation 1) also increases with $z$. Typically, such increases are expressed using power law. However, any suitable function can be used. Let it be $g'(z)$: $M_z = M_{z,0} g'(z)^q$. Since the luminosity is directly proportional to $M$, the net effect is to correct the flux by an additional factor of $g'(z)^{-q}$. This mass dependence of quasars' luminosity is already included in determining the luminosity using the standard concordance model background (e.g., Vestergaard & Osmer 2009, Marziani & Sulentic 2014, Dultzin et al. 2020). This mass dependence must therefore be taken into consideration for the VPC model.

Thus, the modified flux becomes

$$F = \frac{L_s}{4\pi S_k(d_P)^2} \frac{g(z)^{-4} g'(z)^{-q}}{(1+z)^2 g(z)} = \frac{L_v}{4\pi S_k(d_P)^2} \frac{g(z)^{-5} g'(z)^{-q}}{(1+z)^2} \equiv \frac{L_v}{4\pi d_L^2}. \tag{14}$$

The last part of the equation defines the luminosity distance $d_L$. Therefore,

$$d_L = S_k(d_P)(1+z)g(z)^{5/2} g'^{q/2}. \tag{15}$$

We may now write the distance modulus $\mu$

$$\mu \equiv 5\log\left(\frac{d_L}{1\text{Mpc}}\right) + 25 = 5\log\left(\frac{S_k(d_P)}{1\text{Mpc}}\right) + 25 + 5\log(1+z) + 12.5\log g(z) + 5\frac{q}{2}\log g'(z). \tag{16}$$

Before we proceed to fit the observed data, we need to find the expression for the proper distance in the VPC model. Following Gupta (2022b),

$$d_P = \frac{c_0}{H_0} \int_0^z \frac{dz \exp[(1+z)^{-\alpha} - 1]}{E(z)}. \tag{17}$$

Here $H_0$ is the Hubble constant and $E(z)$ is the Peebles function in terms of the scale factor $a \equiv 1/(1+z)$:

$$E(a)^2 = \exp(a^\alpha - 1)\left[\Omega_{m,0} a^{-3}\{1 + \alpha \mathcal{F}(\alpha, a)\} + \Omega_{r,0} a^{-4} + \Omega_{\Lambda,0}\exp(a^\alpha - 1)\right] + \frac{\Omega_{k,0}}{a^2}\exp[2(a^\alpha - 1)]. \tag{18}$$

Here $\Omega_{m,0}$, $\Omega_{r,0}$, and $\Omega_{\Lambda,0}$ are respectively the matter, radiation, and dark-energy densities at present relative to the critical energy density $\varepsilon_{c,0} \equiv 3c_0^2 H_0^2/(8\pi G_0)$, $\Omega_{k,0} = 1 - \Omega_{m,0} - \Omega_{r,0} - \Omega_{\Lambda,0}$ is the relative curvature energy density, and the function $\mathcal{F}(\alpha, a)$ is

$$\mathcal{F}(\alpha, a) \equiv a^3 \exp(a^\alpha - 1)\left[\int_1^a \frac{a'^{\alpha - 4}}{\exp(a'^\alpha - 1)} da'\right] \tag{19}$$

## 4. TYPE xA QUASARS HUBBLE DIAGRAM

Let us consider the Hubble diagram of the xA quasars as studied by Mariziani and collaborators (e.g., Mariziani and Sultenic 2014, Dultzin et al. 2020, Fig. 3). With $d_P$ from equation (17) substituted in equation (16), we can fit the quasar data (Dultzin et al. 2020), which has 253 data points, using the VPC cosmological parameter determined with SNe Ia data fit (Gupta 2022b). However, the parameters in the last term in equation (16) have to be determined by fitting the data. These parameters can then be checked against the same parameters determined by fitting the quasar mass vs. redshift data (e.g., Vestergaard & Osmer 2009). The $z - \mu$ fit can then be compared with the fit with the concordance model ($\alpha = 0, q = 0$).

The most convenient form of the $g'$ function is the same as for the $g$ function. Nevertheless, we will try other functions as well, i.e., the power-law and $g'$ with the same form as for $g$, i.e., $g' = \exp[(1+z)^{-\beta} - 1]$, but not constraining $\beta = 1.8$ as we normally do for $g$. We find the latter to be the best among those we have tried, as discussed later in this paper, and that is what we have used in equation (16) to fit the 253 xA quasars data points.

The fit results for the concordance model with priors $H_0 = 70$ km s$^{-1}$ Mpc$^{-1}$, $\Omega_{m,0} = 0.3$, and $\Omega_{\Lambda,0} = 0.7$, and for the VPC model with priors $H_0 = 70.8$ km s$^{-1}$ Mpc$^{-1}$, $\Omega_{m,0} = 0.271$, and $\Omega_{\Lambda,0} = 0.175$ are shown in Figure 1. Our reason for using priors is to explore how these model parameters obtained from fitting primarily the standard candle SNe Ia data (Gupta 2022b) fit the qusars data. The fit to the 253 data points for the concordance model yields the goodness of fit parameters $\chi^2_{pdf} = 1.80$ and $R^2 = 0.830$, whereas for the VPC model, the fit gives $\chi^2_{pdf} = 1.68$ and $R^2 = 0.843$; $\chi^2_{pdf}$ is $\chi^2$ per degree of freedom. Although the VPC model appears to yield a significantly better fit, it is likely due to the two unconstrained parameters $q$ and $\beta$ in the VPC model. The fit determined these parameters as $q = -7.46 \pm 1.42$ and $\beta = 1.05 \pm 0.33$ which are about the same as determined below, considering their 95% confidence bounds as shown by superscripts and subscripts of the values of respective parameters.

## 5. QUASAR MASS EVOLUTION

We have taken the data of the large bright quasar survey (LBQS) included in the Vestergaard and Osmer (2009) paper to derive parameters of two power-law functions and two $g$ functions. LBQS data is for 978 quasars with their redshift ranging from $z = 0.202$ to $3.36$ and mass ranging from $M_Q = 10^{7.17} M_\odot$ to $10^{10.5} M_\odot$ provided on $\log M_Q$ scale. It covers the range of data we will use for the Hubble diagram from Dultzin et al. (2020). The results are presented in Table 1. Figure 2 shows the data fit corresponding to the best fit function among the three in Table 1. The relevant parameters in the table are $q$ and $\beta$ as they relate to the $g'$ function; $M_{Q,0}$ is irrelevant for scaling purposes as it is implicitly included in the first term of equation (16).

Comparing the $q$ and $\beta$ values corresponding to the minimum $\chi^2_{pdf}$ (= 0.981, i.e., the middle $g'$ function row) in the table with those determined from the $z - \mu$ fit above, we see that they are close and well within the 95% confidence bounds of their respective values. This match establishes that the VPC cosmology is a viable alternative to the concordance cosmology in the context of the present work.

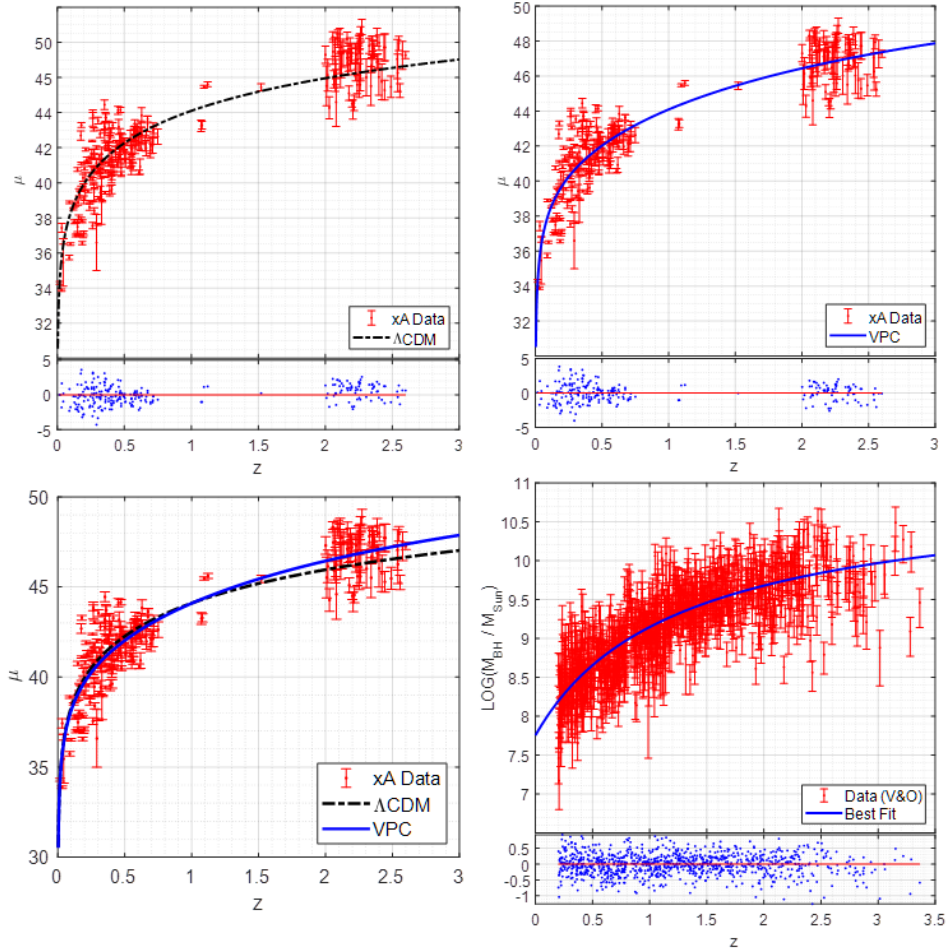

Fig. 1 Hubble diagram of xA quasars fitted with the VPC model and the concordance ΛCDM model and their mass function. Top-left: The VPC model data-fit and the residuals plot. Top-right: The ΛCDM model data-fit and the residuals plot. Bottom-left: Comparison of the two models fitted to the data. The data is as discussed by Dultzin et al. (2020). The 253 data points are separated in sets according to the Hβ and AlIII 1860 virial luminosity estimators: Hβ – 175 points at $z < 0.75$ and 7 points at $z > 1$, and AlIII 1860 – 71 points at $z > 2$. Bottom-right: Mass variation of the quasars with redshift (Vestergaard & Osler 2009). The residual plots are shown as well at the bottom of single model fits.

Table 1 The mass function parameters determined by fitting data (Vestergaard and Osmer 2009) with a power-law and alternative $g'(z)$ functions. The values in subscripts and superscripts represent 95% confidence bounds.

| Function $M_Q = M_{Q,0} g'^q$ | $\log M_{Q,0}$ | $q$ | $\beta$ | $\chi^2_{pdf}$ | $R^2$ |
|---|---|---|---|---|---|
| $g' = z$, i.e., $\log M_Q = \log M_{Q,0} + q \log z$ | $9.166 \pm 0.021$ | $1.634 \pm 0.080$ | NA | 0.908 | 0.622 |
| $g' = \exp[(1+z)^{-\beta} - 1]$, i.e., $\log M_Q = \log M_{Q,0} + \frac{q}{2.3026}[(1+z)^{-\beta} - 1]$ | $7.753 \pm 0.175$ | $-7.838 \pm 1.690$ | $0.7588 \pm 0.3392$ | 0.891 | 0.630 |
| $g' = \exp[(1+z)^{-\beta} - 1]$ with $\beta = 1.8$ | $7.199 \pm 0.102$ | $-6.492 \pm 0.322$ | 1.8 fixed | 0.926 | 0.615 |

## 6. X-RAY AND UV EMISSION CORRELATION

In this section, we will try to study how the X-ray and UV emissions and their relation in quasars evolve with varying physical constants and explore how well the VPC model fits the Hubble diagram for these quasars. (We will label this case as the XUV case to distinguish it from the xA case we have studied above.) The quasar luminosities involved are for the X-ray continuum and the UV spectrum. The black hole accretion disk is the primary producer of UV photons and a corona of hot material forms above the accretion disk. The UV continuum waveband is formed by the thermal emission of the accretion disk. X-ray continuum produced likely from inverse Compton scattering of UV photons produced in the inner part of the hot accretion disk. So, we will need to see how the thermal spectrum and the inverse Compton scattering scale when physical constants vary.

Thermal energy flux is given by (Maoz 2016, p 13)

$$F_{UV} = \frac{8\pi^5 k^4 T^4}{60 c^2 h^3} \sim \frac{g^8 T}{g^5} \sim g^3 T, \quad (20)$$

where $T$ is the temperature of the thermal emission region. Equation (20) can be considered for the scaling of the thermal UV radiation. The inverse Compton scattering energy boost $\gamma$ for scaling purpose may be written as (Maoz 2016, p 223)

$$\gamma = \frac{kT_e}{m_e c^2} n_e \sigma_T x \sim \frac{g^2}{g^2} g^{-3} g T_e \sim g^{-2}, \quad (21)$$

where $T_e$ is the corona temperature and $n_e$ is its electron density, and $x$ represents the corona size. Recall that electron density scales as $g^{-3}$ since volume is dimensionally cube of length and length scales as $g$. Since the X-ray continuum is created by boosting thermal UV photons, we get by combining equations (20) and (21)

$$F_X \sim g^3 g^{-2} \sim g. \quad (22)$$

We should now consider the absorption of the radiation in the matter surrounding the accretion disk as well as in the interstellar medium. The absorption will be proportional to density, which scales as $g^{-3}$. As a result, the X-ray flux reduction will scale as $g^3$, i.e., $F_X \sim g^4$. We have not yet considered the absorption in the interstellar medium, which is known to absorb higher energy photons preferably and is a function of distance traveled. Now distance scales as $g$ in VPC. Therefore, we may write the X-ray flux scaling as

$$F_X \sim g^5. \quad (23)$$

From equations (20) and (23), we have

$$\log(F_{UV}) - \log(F_X) = C - 2\log(g). \quad (24)$$

Here $C$ is a positive constant from observations (Lusso et al. 2020). It would mean that in our model, when $g$ is decreasing with increasing redshift $z$, one expects $\log(F_{UV}) - \log(F_X)$ to *increase* with increasing $z$ as expected (e.g., Reeves & Turner 2000, Goulding et al. 2018). This is confirmed by fitting the observational data for 2421 quasars (Lusso et al. 2020). We found $\log(F_{UV}) - \log(F_X)$ to *increase* with $z$ as follows (Fig. 2):

$$\log(F_{UV}) - \log(F_X) = 3.2154 - 2.0709 \log(g). \quad (25)$$

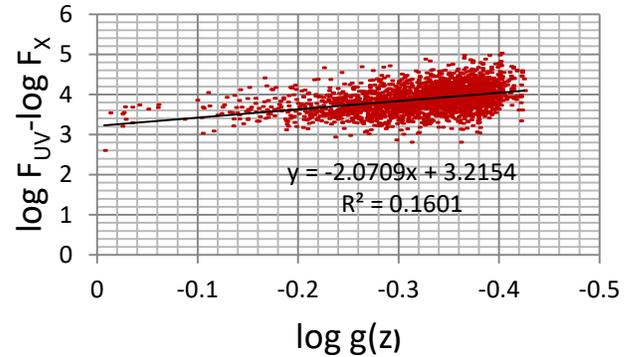

Fig. 2 Plot of $\log F_{UV} - \log F_X$ against $\log g(z)$ and the trend line fitting the data (Lusso et al. 2020). Note: $\log g(0) = 0$ and $\log g(7.5) = -0.425$.

Equations (24) and (25) are essentially the same, confirming our consideration in arriving at equation (24).

## 7. XUV QUASARS' HUBBLE DIAGRAM

The scaling of the quasar luminosity we have used for xA quasars is probably not applicable for XUV class of quasars because the latter does not satisfy the criterion presented in Sec. 2. Equation (1) remains the same but $\dot{M}$ may no longer be defined by equation (8), i.e., by the Eddington limit. Let us therefore treat $\dot{M}$ scaling to be unknown, i.e., $\dot{M} \sim g(z)^p$, and see if $p = 2$, which we have used for the xA quasars, makes sense for the XUV quasars by fitting the Hubble data (2421 quasars) for the latter (Lusso et al. 2020). Thus, we have $L \sim g(z)^{p+2}$, i.e., $L_v = L_s g(z)^{p+2}$, i.e., $L_s = L_v g(z)^{-p-2}$. Thus, equation (16) becomes

$$\mu = 5\log\left(\frac{S_k(d_P)}{1\text{Mpc}}\right) + 25 + 5\log(1+z) + 5\frac{p+3}{2}\log g(z) + 5\frac{q}{2}\log g'(z). \quad (16)$$

The values of $q$ and parameter $\beta$ in function $g' = \exp[(1+z)^{-\beta} - 1]$ will be different from those for the xA quasars. It is because the mass evolution for XUV quasars is not expected to be the same as for xA quasars. These parameters can be determined by fitting $z - \mu$ data and would relate to the evolution of black hole mass for XUV class of quasars.

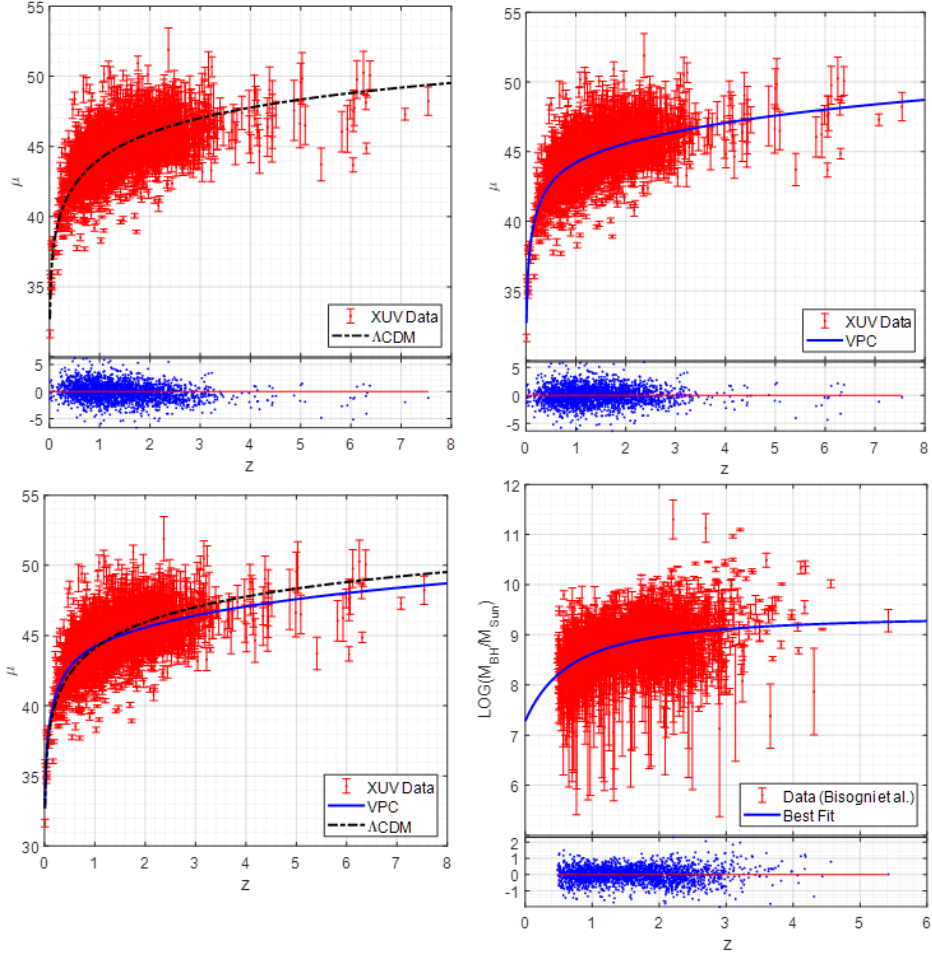

Fig. 3 Hubble diagram of XUV quasars fitted with the VPC model and the concordance ΛCDM model and their mass function. Top-left: The VPC model data-fit and the residuals plot. Top-right: The ΛCDM model data-fit and the residuals plot. Bottom-left: Comparison of the two models fitted to the data (Lusso et al. 2020). Bottom-right: Mass variation of the quasars with redshift (Bisogni et al. 2021). The residual plots are shown as well at the bottom of single model fits.

The results are presented in Fig. 3. Using the same VPC priors as for the xA quasars, i.e., $H_0 = 70.8 \text{ km s}^{-1} \text{ Mpc}^{-1}$, $\Omega_{m,0} = 0.271$, $\Omega_{\Lambda,0} = 0.175$, and $p = 2$, we get $q = -4.43 \pm 0.12$, and $\beta = 3.23 \pm 0.37$ with $\chi^2_{pdf} = 2.47$ and $R^2 = 0.431$.

These numbers may be compared with the concordance model goodness-of-fit parameters $\chi^2_{pdf} = 2.59$ and $R^2 = 0.402$. The goodness-of-fit parameters for the VPC model are better than for the concordance model. This may be expected, as the former has two free parameters, whereas the latter has none. If we make $p$ also as a free parameter, we get unacceptably high and low values for the three parameters with 95% confidence bounds, indicating very high degeneracy among them ($q = -10.0 \pm 29.8$, $\beta = 2.40 \pm 1.89$ and $p = 7.79 \pm 30.44$) with essentialy no improvement in the goodness-of-fit parameters ($\chi^2_{pdf} = 2.47$ and $R^2 = 0.431$). Thus, it appears prudent to constrain $p = 2$, i.e., the same as for xA quasars, i.e., take $\dot{M} \sim g(z)^2$ from equation (8) also for XUV quasars. This may be interpreted that while $\dot{M}$ may not be given by equation (8), it may still be *proportional* to $8\pi rcm_p/\sigma_T$, which leads to $\dot{M} \sim g^2$.

The observed mass evolution of the XUV quasar can be determined by fitting the corresponding black hole mass data (Bisgoni et al. 2021, Bisgoni – private communication) with the function $M_Q = M_{Q,0} g'^q$ with $g'(z) = \exp[(1+z)^{-\beta} - 1]$. The procedure is the same as for the case of xA quasars. The fit is shown in Fig. 3 (bottom-right). We got $q = -4.89 \pm 0.61$ and $\beta = 1.44 \pm 0.64$. While $q$ is relatively close to the value determined from fitting the Hubble diagram, $\beta$ value is about half. This may not be surprising considering large scatter and error bars in the observed black hole masses. Alternatively, it may indicate that the VPC mass evolution formula is not suitable for the XUV quasars.

Table 2 Constraints on $(\dot{G}/G)_0$ determined by various methods. CL means 'confidence limit', and NA means 'not available'. Here Min. $\leq (\dot{G}/G)_0 \leq$ Max.

| Look back time in years | Method | $(\dot{G}/G)_0$ in units of $10^{-12}$ yr$^{-1}$ | | | Reference |
|---|---|---|---|---|---|
| | | Max. | Min. | CL | |
| ~40 | Helioseismology | <0.2 | | 95% | Bonanno & Fröhlich (2020) |
| ~45 | Lunar Laser Ranging | 0.147 | -0.005 | NA | Hofmann & Müller (2018) |
| ~45 | Planetary Ephemeris | 0.078 | -0.07 | 95% | Pitjeva & Pitjev (2013) |
| ~60 | White Dwarf Pulsations | 40 | -250 | 95% | Benvenuto et al (2004) |
| ~3000 | Pulsar Timing | 9 | -27 | 68% | Kaspi et al (1994) |
| ~100 Million | Pulsar Mass | 3.6 | -4.8 | 95% | Thorsett (1996) |
| ~200 Million | Gravitational Waves | 20000 | -4000 | 90% | Vijaykumar et al. (2021) |
| ~4 Billion | Young Sun Luminosity | | -4 | NA | Sahini & Shtanov (2014) |
| ~5 Billion | Supernovae Type Ia | 10 | NA | NA | Gaztanaga et al (2001) |
| ~5 Billion | White Dwarf Cooling | | <-50 | NA | Althaus et al. (2011) |
| ~10 Billion | Stellar Astroseismology | <=5.6 | | 95% | Bellinger & C.-Dalsgaard (2019) |
| ~10 Billion | Age of Globular Clusters | 7 | -35 | NA | Degl'Innocenti (1996) |
| ~14 Billion | Cosmic Microwave Background | 1.05 | -1.75 | 95% | Wu & Chen (2010) |
| ~14 Billion | Big Bang Nucleosynthesis | 4.5 | -3.6 | 95% | Alvey et al. (2020) |
| ~14 Billion | VPC – Quasar Hubble Diagram | 394 | 386 | 95% | This paper |

## 8. DISCUSSION

The advantage of considering quasars for measuring cosmological distances and testing cosmological models as compared to the type Ia supernovae is that quasar greatly extends the cosmological time scale: from the cosmic time corresponding to $z \sim 1.5$ for supernovae to $z > 7.5$ for quasars. As pointed out by Mariziani and Sulentic (2014), the differences in the redshift coverage mean that while supernovae cover the epoch of $\Lambda$ dominance, quasars includes the epoch when $\Omega_M$ ruled the expansion of the Universe. However, currently, the statistical errors and scatter in the quasar data are too large compared to the supernovae data to consider the former as a reliable standard candle for distance measurement. Therefore, one uses the cosmological parameters determined from SNeIa data to see how well these parameters fit the quasar data statistically.

In their recent paper, Lusso et al. (2020) wrote: 'We confirm that, while the Hubble diagram of quasars is well reproduced by a standard flat $\Lambda$CDM model (with $\Omega_M = 0.3$) up to $z \sim 1.5$, ...., a statistically significant deviation emerges at higher redshifts, in agreement with our previous works (e.g., Risaliti & Lusso 2015, 2019; Lusso et al. 2019) and other works on the same topic (e.g., Di Valentino et al. 2020).' They then tried to fit the quasar Hubble diagram with a flat $w_0 w_a$CDM model, which is a commonly used extension of the standard $\Lambda$CDM model, where the parameter $w$ of the equation of state of the dark energy is considered to vary with redshift $z$ according to the parametrization $w(z) = w_0 + w_a \times (1 - a)$, with $a = (1 + z)^{-1}$ being the scale factor. It essentially introduces two extra parameters, $w_0$ and $w_a$ for fitting the Hubble diagram data without cross-checking for their correctness or validity.

In the VPC approach, we basically have two parameters ($q$ and $\beta$), other than those determined previously by fitting the Pantheon SNeIa data, to fit the quasar data. However, we have tried to validate these parameters from the observed variation of the quasar black hole masses with redshift. We see that for the case of xA quasars, the parameters $q = -7.46 \pm 1.42$ and $\beta = 1.05 \pm 0.33$ determined from fitting their Hubble diagram, and $q = -7.84 \pm 1.69$ and $\beta = 0.759 \pm 0.339$ determined from fitting the quasar mass data, match within their 95% confidence bounds. However, for the XUV quasars, we get $q = -4.43 \pm 0.12$, and $\beta = 3.23 \pm 0.37$ from fitting their Hubble diagram and $q = -4.89 \pm 0.61$ and $\beta = 1.44 \pm 0.64$ from fitting quasar mass data: While $q$ match is good, it is not so for $\beta$. The discrepancy may be due to large scatter and errors in the quasar masses or inadequate mass evolution formula used for fitting the Hubble diagram for XUV quasars.

One would notice that there are significant uncertainties in the parameters determined in this study. It may be expected considering that there is a significant scatter in the quasar data.

A bonus feature of this study is that a correlation between the X-ray and UV fluxes exists through the scaling of the two fluxes in the VPC model: We showed that the X-ray flux scales as $F_X \sim g(z)^5$ and the UV flux scales as $F_{UV} \sim g(z)^3$. This leads to $\log(F_{UV}) - \log(F_X) = C - 2\log(g)$ with $C$ being a constant. When we fitted the observed data, we got $\log(F_{UV}) - \log(F_X) = 3.2154 - 2.0709 \log(g)$. This finding bolstered our confidence in the VPC model.

Thus, equation (6) for the function $g = \exp[(1 + z)^{-1.8} - 1] = \exp(a^{1.8} - 1)$ can be considered to reasonably represent the coupling constants' variation given by equation (4). This variation then leads to $\dot{G}/G = 5.4 a^{1.8} \dot{a}/a = 5.4 a^{1.8} H$. At the current time (i.e., $z = 0$ or $a = 1$), $(\dot{G}/G)_0 = 5.4 H_0$, and since $H_0 = 70.8 (\pm 0.7)$ km s$^{-1}$ Mpc$^{-1}$ = $7.23 (\pm 0.07) \times 10^{-11}$ yr$^{-1}$, we get $(\dot{G}/G)_0 = 3.90 (\pm 0.04) \times 10^{-10}$ yr$^{-1}$. The confidence bound for this value is the same as for $H_0$, i.e., 95%. This constraint on $(\dot{G}/G)_0$ is shown in Table 2 along with some others determined by various methods. One immediately notices that the constraint determined by the VPC approach is among the most relaxed. It is primarily due to the concurrent variation of several coupling constants treated in the VPC approach. All other methods ignore the potential variation of constants other than the gravitational constant. In the cases where the evolution of stellar objects is involved on cosmological time scales (e.g., Teller 1948, Degl'Innocenti et al. 1996, Thorsett 1996, Corsico et al. 2013, Bellinger & Christensen-Dalsgaard 2019), there is an additional concern related to energy conservation. It is because energy is not conserved in cosmological evolution and general relativity (Harrison 1981, 1995, Peebles 1993, Baryshev 2008, Velten & Caramês 2021, Gupta 2022a).

Since variations of $c$, $G$, $h$, and $k$ are interrelated as $G \sim c^3 \sim h^3 \sim k^{3/2}$, the constraint we have determined for $G$ applies also to $c$, $h$, and $k$ through $(\dot{c}/c)_0 = (\dot{h}/h)_0 = 1/2 (\dot{k}/k)_0 = 1/3 (\dot{G}/G)_0$. We have recently proposed (Gupta 2022c) how it might be possible to test this relationship among constants in a laboratory using the same equipment, the Kibble balance (e.g., Kibble et al. 1990), that is used for determining Planck constant and calibrating weighing standards with a precision of a few parts per billion (Schlamminger & Haddad 2019).

## 9. CONCLUSION

Astrophysicists have been concerned about the inability of the concordance model to fit the Hubble diagrams of xA and XUV quasars. The variable physical constant approach fits the diagrams satisfactorily using the same constraint on the variation of the coupling constants, i.e., on the variation of the speed of light $c$, the gravitational constant $G$, the Planck constant $h$, and the Boltzmann constant $k$, as the Hubble diagram of SNeIa from the Pantheon data. Since the variation of coupling constants is interrelated in the VPC model through $G \sim c^3 \sim h^3 \sim k^{3/2}$, we can express the general constraint in terms of the Hubble constant $H_0$ as $(\dot{G}/G)_0 = 3(\dot{c}/c)_0 = 3(\dot{h}/h)_0 = 3/2 (\dot{k}/k)_0 = 5.4 H_0 = 3.90(\pm 0.04) \times 10^{-10}$ yr$^{-1}$.

## Acknowledgments

The author is grateful to Dr. Paola Marziani, Dr. Guido Risaliti, and Dr. Susanna Bisogni for providing the data used in this work. He acknowledges an unconditional grant from Macronix Research Corporation in support of the research. Special thanks are due to the editor and reviewer for suggesting improvements to the paper.

## Data Availability

All the data used in this research is available from the cited references.

## Conflict of Interest

This work is performed without any conflict of interest.